\documentclass[%
 reprint,
 amsmath,amssymb,
 aps,
]{revtex4-1}

\usepackage{enumerate}
\usepackage{amsmath}
\usepackage{graphicx}
\usepackage{dcolumn}
\usepackage{bm}

\usepackage{xcolor}

\begin{document}

\preprint{APS/123-QED}

\title{Curved edges in the vertex model increase tissue fluidity}

\author{Michael F. Staddon}
\affiliation{Max Planck Institute of Molecular Cell Biology and Genetics, Dresden, Germany}
\affiliation{Center for Systems Biology Dresden, Dresden, Germany}
\affiliation{Cluster of Excellence, Physics of Life, TU Dresden, 01307 Dresden, Germany}

\author{Carl D. Modes}
\affiliation{Max Planck Institute of Molecular Cell Biology and Genetics, Dresden, Germany}
\affiliation{Center for Systems Biology Dresden, Dresden, Germany}
\affiliation{Cluster of Excellence, Physics of Life, TU Dresden, 01307 Dresden, Germany}

\begin{abstract}
    The Vertex Model for epithelia models the apical surface of the tissue by a tiling, with polygons representing cells and edges representing cell-cell junctions. The mechanics are described by an energy governed by deviations from a target area and perimeter for each cell. It has been shown that the target perimeter, $p_0$, governs a solid-to-fluid phase transition: when the target perimeter is low there is an energy barrier to rearrangement, and when it is high cells may rearrange for free and the tissue can flow like a liquid. One simplification often made is modelling junctions using straight edges. However, the Young-Laplace equation states that interfaces should be circular arcs, with the curvature being equal to the pressure difference between the neighbouring cells divided by the interfacial tension. Here, we investigate how including curved edges alters the mechanical properties of the vertex model and equilibrium shape of individual cells. Importantly, we show how curved edges shift the solid-to-fluid transition point, from $p_0 = 3.81$ to $p_0 = 3.73$, allowing tissues to fluidise sooner than in the traditional model with straight edges.
\end{abstract}

\maketitle

In many biological processes, such as wound healing~\cite{brugues2014forces, ajeti2019wound, tetley2019tissue}, morphogenesis~\cite{bertet2004myosin, kim2021embryonic, maniou2021hindbrain}, and cancer progression~\cite{friedl2009collective, arwert2012epithelial}, epithelial tissues undergo dramatic and irreversible shape changes. For a tissue to change its shape or relax under stresses, individual cells can alter their shape~\cite{etournay2015interplay, clement2017viscoelastic}, for example by anisotropic contractility, cells may divide or die~\cite{ranft2010fluidization}, or cells can rearrange~\cite{tetley2018same}, most famously demonstrated in convergent extension~\cite{keller2000mechanisms, irvine1994cell}. When cells may rearrange easily the tissue is said to be fluid-like and can flow under shears and easily change its shape~\cite{guillot2013mechanics, mongera2018fluid}. The fluidity of the tissue is governed by both subcellular properties, such as contractility or elasticity, and external ones, such as cell-cell or cell-substrate adhesion, and can actively adapt based on environmental cues. Thus, understanding how it is governed remains an important question.

Various theoretical models have been developed to understand this process, including the vertex model~\cite{nagai2001dynamic, farhadifar2007influence, fletcher2014vertex, bi2015density, alt2017vertex}. Originally developed from models of soap films~\cite{nagai2001dynamic}, the vertex model describes the apical surface of an epithelial tissue as a tiling of polygonal cells, with polygons representing cells, edges the cell-cell interfaces, and vertices the tri-cellular junctions. The molecular mechanisms within the cell, such as acto-myosin contractility, cell cortex rigidity, and cell-cell adhesion, are coarse-grained and the mechanics of each cell is described as having an area and perimeter elasticity, and interfacial tension. The vertex positions are then determined by force balance, during which cell rearrangements, also known as T1 transitions or intercalations, may occur. The vertex model, sometimes with additional features such as cell motility, division, or biochemical reactions, has been successful in capturing a range of biological processes such as wound healing~\cite{tetley2018same, ajeti2019wound}, tissue dynamics~\cite{maniou2021hindbrain, mao2013differential, yousafzai2022cell}, cell sorting~\cite{perrone2016non}, and more.

Despite its simplicity, the vertex model exhibits a rich variety of mechanics~\cite{farhadifar2007influence, bi2015density, duclut2021nonlinear, tong2022linear, hernandez2022anomalous, staddon2023role, hernandez2023finite}. An important prediction of the vertex model is the solid-to-fluid phase transition~\cite{bi2015density}, which is controlled by one parameter, $p_0$, the preferred shape index of the cell. When $p_0 < p_* \approx 3.81$, there exists a finite energy barrier to rearrangements, while for $p_0 > p_*$ rearrangements occur for free, and such trends have been observed in experimental data. However, one feature that was lost when adapting the model from soap film models was the use of curved edges in the form of circular arcs, as predicted by the Young-Laplace Law, which states that the curvature between two interfaces is equal to the pressure difference divided by the interfacial tension. 


However, so far there has been no studies on how curved edges affects the fluidity of the tissue, despite some research demonstrating the importance of curved edges in the mechanics of the vertex model~\cite{ishimoto2014bubbly, perrone2016non, schaumann2018force, kim2021embryonic, kim2024two}, or in soft particle models~\cite{boromand2018jamming, treado2021bridging, vetter2024polyhoop}, during invasion, migration, or jamming, using either circular arcs or subdivided straight edges. For example, Perrone et al~\cite{perrone2016non} found that curved cell edges were important in controlling tissue mechanics during simulations of invasion and engulfment. This extra degree of freedom could also allow the tissue to further relax energy under a T1 transition and could therefore give different results for the mechanical response of the tissue.

Thus, we ask how the curved-edge model and straight-edge model differ in their predictions of tissue jamming. First, we model a simple 4-cell cluster undergoing a T1 transition and show how the solid-to-fluid transition point is lower when curved edges are allowed. Next, we increase the size of the tissue and show how this transition point converges to $p_0 = 3.729$, which is lower than in the straight-edge model. Finally, we investigate a more realistic disordered tissue and show how the the energy barrier is consistently lower in the curved-edge vertex model. Additionally, while the linear shear modulus does not differ between the straight and curved-edge models, the energy is lower under large shears with curved-edges as T1 transitions occur sooner.


The mechanics of the tissue are described from deviations in cell shape from a target area, $A_0$, and perimeter $P_0$. Assuming uniform cell properties, the non-dimensional energy takes the form
\begin{equation}
    E = \frac{1}{2} \sum_\alpha (a - 1)^2 + \frac{1}{2} \sum_\alpha \Gamma (p - p_0)^2.
\end{equation}
where $\alpha$ indicates a sum over all cells, $a$ is the non-dimensional area with target area $1$, $\Gamma$ the contractility or perimeter elasticity, and $p$ is the dimensionless perimeter with target shape index $p_0$. The first term represents volume conservation of the cell in 3 dimensions, resulting in a 2 dimensional area elasticity. The second term represents perimeter elasticity, which is a combination of cytoskeletal elasticity, cortical tension, and cell-cell adhesion. The target shape index $p_0$ can be either positive or negative, depending on the balance of forces, with tension reducing the value and adhesion increasing it. When allowing for curved edges, each edge is described by a circular arc with curvature $K_{ij}$, for the edge connecting vertices $i$ and $j$.

\begin{figure}[h]
\includegraphics[width=\columnwidth]{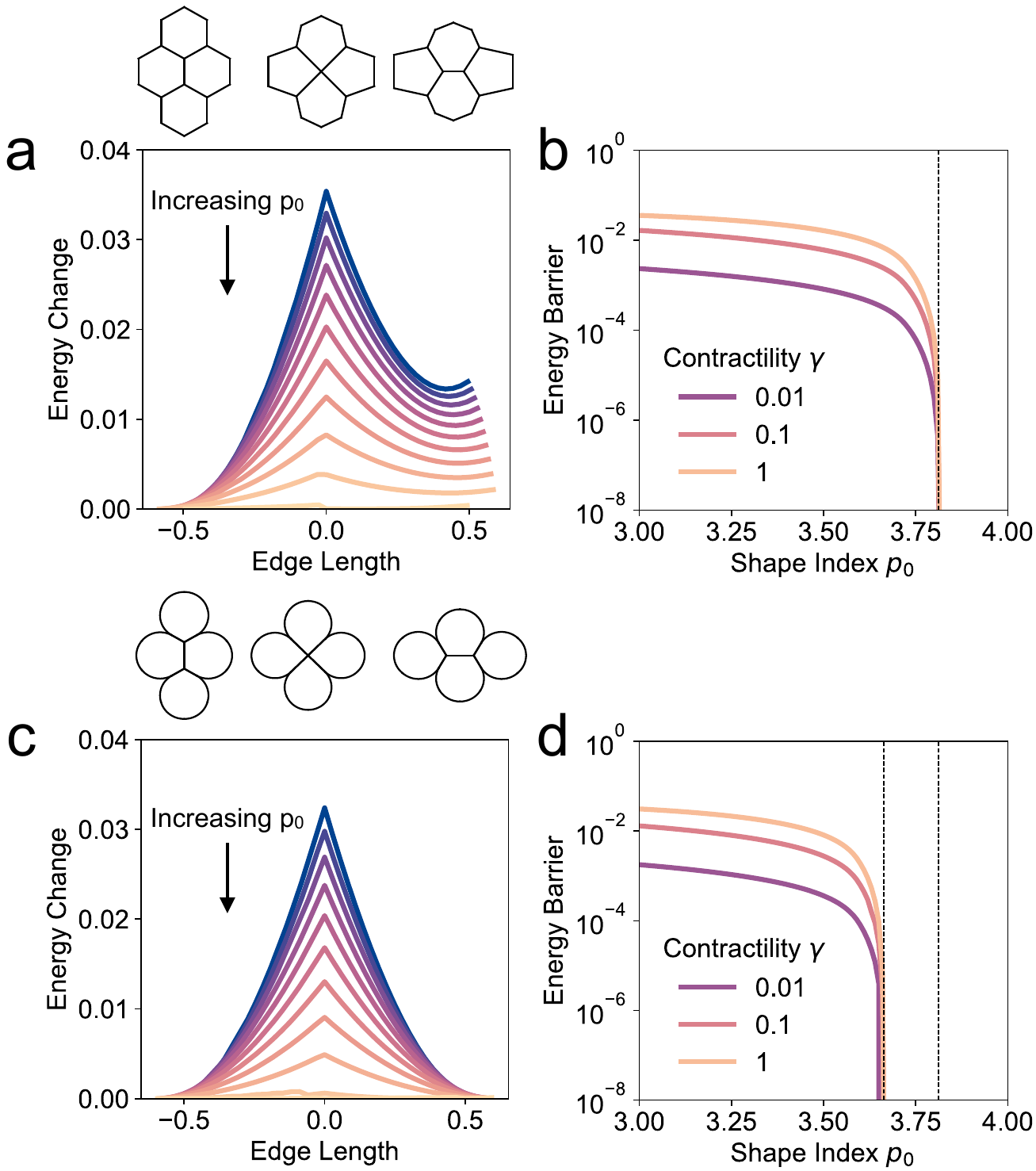}
\caption{Curved edges reduce the solid-to-fluid transition point. (a) Energy change during a T1 transition for straight edges. Brighter colours indicate higher $p_0$. Negative length means before the T1, positive length means after the T1. $p_0$ ranges from $1.8$ to $3.8$. (b) Energy barrier from a T1 as a function of $p_0$. For $p_0 > 3.81$ there is no energy barrier. (c) Energy change during a T1 transition for curved edges. (d) Energy barrier for curved edge vertex model against $p_0$. For $p_0 > 3.66$.}
\label{fig:1}
\end{figure}

The force on the vertex $i$ with position $\mathbf{x}_i$ is defined as
\begin{equation}
    \mathbf{F}_i = -\frac{\partial E}{\partial \mathbf{x}_i}
\end{equation}
such that the force acts to minimises the tissue energy. As vertices move, the tissue may undergo a T1 transition, also known as a rearrangement or intercalation, in which an edge will shrink to a point, and a new edge is formed perpendicular to it connecting two initially unconnected cells (Fig.~\ref{fig:1}a). When a T1 transition can occur with no energy barrier, the tissue is said to be in the fluid phase, otherwise it is in the solid phase.

Throughout the rest of the paper, we use a conjugate-gradient descent method to find the mechanical equilibrium, using all free variables, including vertices, edge curvature, and periodic boundaries where applicable, unless otherwise fixed. T1 transitions occur when edges shrink below a length threshold, with the threshold chosen to be small so that the choice does not affect the mechanics of the system. These models are implemented using Surface Evolver~\cite{brakke1992surface}.


We first consider a 4 cell cluster with free boundaries that undergoes a T1 transition. When cells are constrained to have straight edges, we start from the ground state of 4 cells with 6 edges each (Fig.~\ref{fig:1}a). Next, we shrink the edge until it becomes a point and then perform a T1 transition, in which the two cells which were previously not connected gain an interface, while the previously two connected cells lose theirs. We then increase the length of this new junction perpendicularly to the original. As the edge shrinks, the energy of the cells increases and becomes a maximum once the edge has zero length (Fig.~\ref{fig:1}a). After the T1, the energy decreases as the edge length increases, but the new arrangement of the cells has a higher configuration compared to before the T1 as we now have two pentagonal cells and two heptagonal cells, compared to four hexagonal ones.

As the preferred shape index $p_0$ is increases, the energy barrier required to perform a T1 decreases until the shape index of a regular pentagon at $p_0 \approx 3.813$ (Fig.~\ref{fig:1}b). The highest energy point is when the junction has zero length, giving two pentagonal cells. Thus, when they can achieve both their target area and perimeter there is no energy penalty. For $p_0 > 3.813$ there is no energy barrier and the tissue is said to be in a fluid state.

Next, we repeat this simulation but allow edges to be circular arcs, as per the Young-Laplace law (Fig.~\ref{fig:1}c). In this instance, the energy barrier is lower than for the straight edges, (Fig.~\ref{fig:1}d) and the solid-to-fluid transition happens earlier, at $p_0 \approx 3.66$, since the cells are more circular and can have a reduced perimeter for the same area relative to pentagonal cells (Fig.~\ref{fig:1}c). More over, the energy barrier is symmetric about the T1 transition since the cells are not constrained to be pentagonal or heptagonal. While this is a simple toy example, it shows how the constraints and assumptions put into a model, in this case only allowing straight edges, can significantly change the behaviour of the model.

We can calculate analytically the transition point in this 4-cell model when curved edges are allowed. The highest energy occurs when the central edge has zero length. Since all cells can be considered identical, due to the curved edges, we consider the shape of a single cell that minimises energy. The only requirement is that the boundary passes through the central point, and so technically the energy minimum can be for a circular cell, ie $p_* = 2 \pi / \sqrt{\pi}$. However, this leads to large overlaps of cell areas. Instead, we consider the additional requirement that cells may not overlap. The cell-cell interfaces must be straight lines, due to the balance of pressure, while the outer edges are circular arcs. To minimise the shape index of the cell we want to maximise how much is contained in the circular section, meaning the arc is tangent to the straight line at the vertex, as can be seen in Fig.~\ref{fig:1}c. It can be shown that the interfacial length $L$ equal to the radius of the circular arc, since the two cell-cell junctions and lines connecting the vertices to the circle center form a square.


The perimeter of such as shape is given by
\begin{equation}
    p = \left(2 + \frac{3 \pi}{2} \right) L
\end{equation}
and area
\begin{equation}
    a = \left(1 + \frac{3 \pi}{4}\right) L ^2.
\end{equation}
Since we have unit area we calculate the minimum perimeter as
\begin{equation}
    p_* = \sqrt{4 + 3 \pi} \approx 3.664
\end{equation}
which matches the numerical simulations. Thus, in a 4-cell cluster the solid-to-fluid transition point has shifted from $p_0 = 3.814$ when edges are forced to be straight, to $p_0 = 3.664$ when they are allowed to curve.

\begin{figure}[h]
\includegraphics[width=\columnwidth]{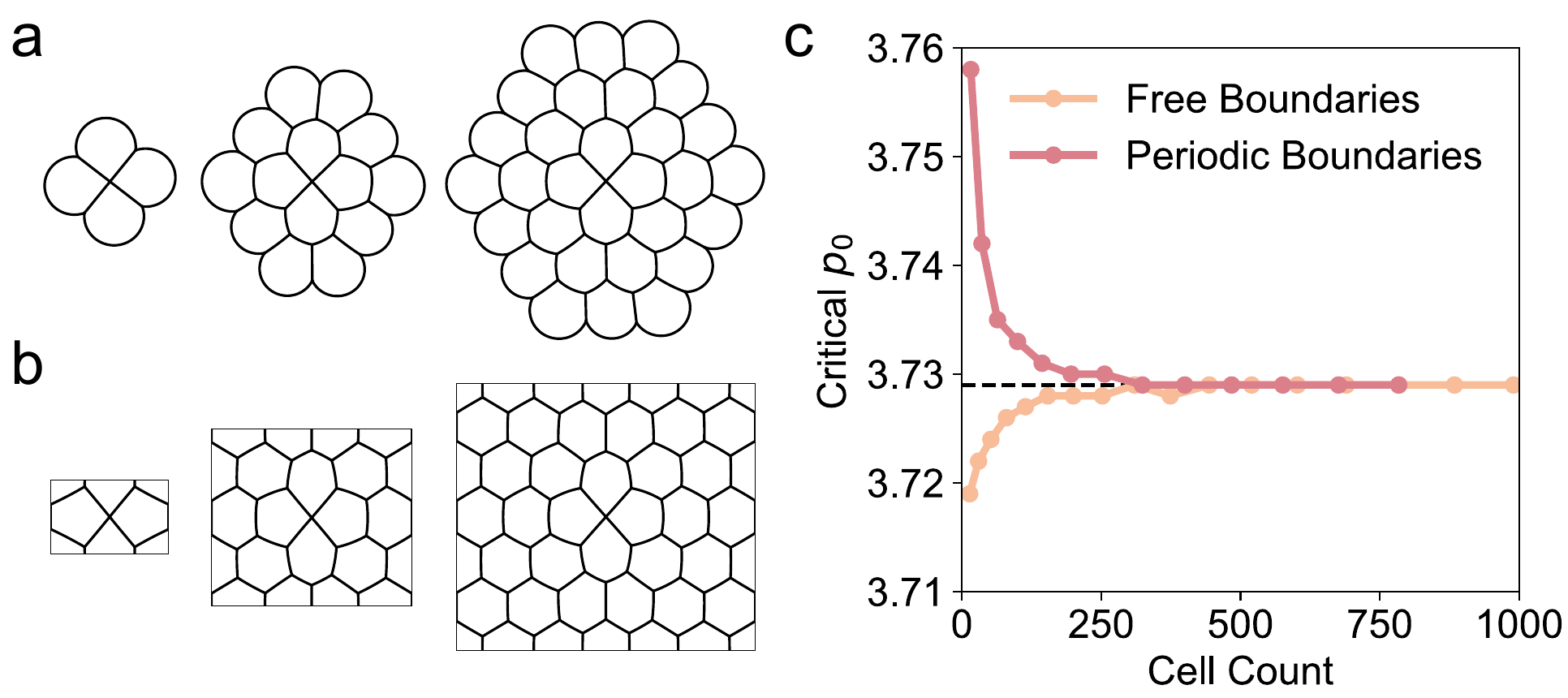}
\caption{The transition point in large tissues converges to $p_* = 3.729$. (a) Examples of tissue shapes for free boundaries when the middle edge is collapsed, with $p_0$ chosen as the critical value. (b) Examples of tissue shape used with periodic boundaries. (c) The critical $p_0$, where the energy barrier becomes zero, against cell count for free and periodic boundaries. The dashed line indicates $p_0 = 3.729$.}
\label{fig:2}
\end{figure}

For a more realistic model, we consider larger tissues with both periodic and free boundaries. We use a hexagonal lattice as the ground state and measure the energy required to shrink an edge at the center of a tissue. For the free boundary tissue we generate concentric rings of cells about the 4-cell cluster (Fig.~\ref{fig:2}a). For periodic boundaries we use a lattice of $n$ by $n$ cells where $n$ is even (Fig.~\ref{fig:2}b). For both curved-edge and straight-edge models with periodic boundaries, the ground state is the same lattice of regular hexagonal cells. Despite this, we observe a significant shift in critical $p_0$ due to edge curvature which appears during a T1 transition.

As the number of cells increase we see the critical $p_0$ for free boundaries increasing, since we have relatively more constraints on the tissue, and for periodic boundaries we see the critical $p_0$ decreasing (Fig.~\ref{fig:2}c). In both cases, as the cell number increases we see convergence of the critical $p_0 \approx 3.729$, at around 400 cells. Thus, allowing for curved edges in the vertex model shifts the solid-to-fluid transition point from $p_* \approx 3.813$ to $p_* \approx 3.729$. This is almost equivalent to shifting the fluidity transition by a whole polygon class, from the shape index of a pentagon to near the shape index of a regular hexagon $p_0 \approx 3.722$. With straight edges, cells must be able to form regular pentagons to rearrange with no energy cost, while for curved edges cells which have a target shape index slightly higher than a regular hexagon can already fluidise.

Next, we consider the case of a disordered tissue. Starting from a periodic lattice of 4 cells, we grow the tissue to 100 cells using a sequence of random cell divisions followed by energy minimisation, as described in Farhadifar et al (cite). In this section we first quantify the curvature seen in disordered tissues, and then compare the mechanical response to the straight-edge vertex model.

\begin{figure}[h]
\includegraphics[width=\columnwidth]{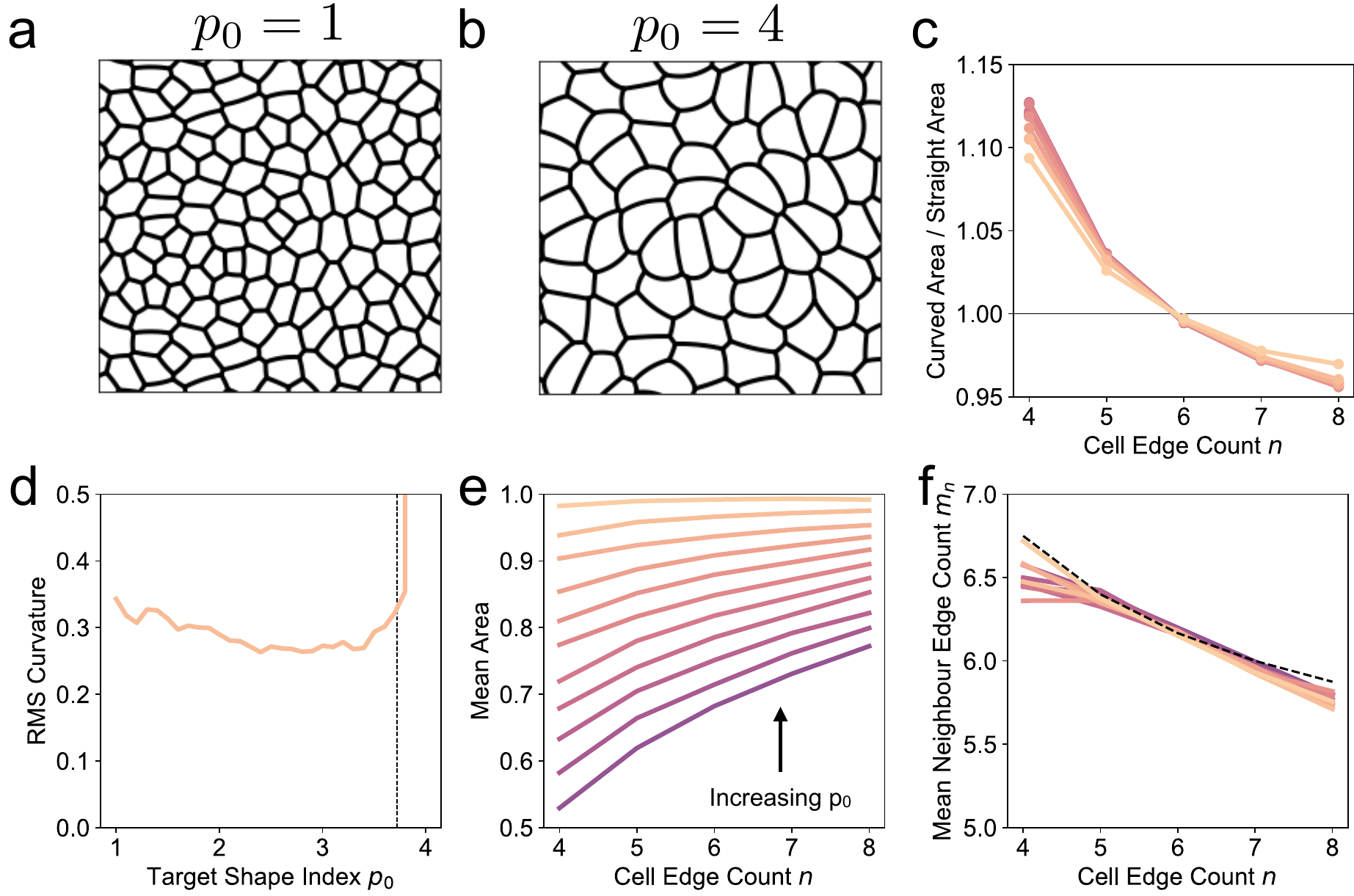}
\caption{Cell shape quantification in disordered tissues (a) An example of a tissue in the solid regime, $p_0 = 1.5$. (b) An example of a tissue in the fluid regime, $p_0 = 4.0$. (c) Mean curved area over straight area for different cell edge counts, for $p_0 = 1.8$ (dark) to $p_0 = 3.8$ (light). (d) Root mean square curvature against $p_0$. (e) Mean cell area against polygon number, for $p_0 = 1.8$ (dark) to $p_0 = 3.8$ (light). (f) Mean neighbour edge count against cell edge count, for $p_0 = 1.8$ (dark) to $p_0 = 3.8$ (light). The dash line indicates the prediction from the Aboav-Weaire Law.}
\label{fig:3}
\end{figure}

Disordered tissues show a variety of cell areas, perimeters, and neighbour count. As a result, the difference in pressure and tension gives curved interfaces between cells. For tissues in the incompatible regime, $p_0 < p_6$, we typically observe smaller cells having edges which curve outwards, while larger cells have edges which curve inwards (Fig.~\ref{fig:3}a). Intuitively, this is because smaller cell are more compressed and have a higher pressure compared to larger cells, thus by Laplace's law the edge will curve into the larger cell. For tissues in the compatible regime, cells attain both the preferred area and perimeter and are in a degenerate ground state. We observe large amounts of curvature on all the edges (Fig.~\ref{fig:3}b). However, a model with straight edges could also satisfy the area and perimeter constraints, for example by elongating the cells. 

Cells with fewer neighbours are on average smaller, as famously observed in biological systems by Lewis's law, and so tend to curve outwards more often. To quantify this, we compare the area of the cell with curved edges, the "curved area", to the area of the cell if there was no curvature, the "straight area" (Fig.~\ref{fig:3}c). Cells with only 4 neighbours are on average 10\% larger when using curved area compared to straight area. As the number of neighbours increase the ratio of areas decreases, with 6-sided cells having an equal curved area as straight, suggesting that on average they have zero curvature, while larger cells can have 10\% smaller curved area compared to straight area.

Cell shape also depends on the shape index $p_0$. There is a non-monotonic relationship between mean squared curvature and the target shape index (Fig.~\ref{fig:3}d). As $p_0$ increases, the curvature decreases and then rapidly increases as $p_0$ approaches $p_6$. We can understand this by considering the variability in cell sizes as a function of $p_0$. For low $p_0$, we observe Lewis's law; cells with fewer edges have much smaller area than high polygon count cells, meaning there is a larger pressure difference between them (Fig.~\ref{fig:3}e). This difference in size decreases as $p_0$ increases, and for $p_0 > p_6$ many take their target shape and have zero pressure, and so can easily bend, resulting in a high curvature. Additionally, for all $p_0$ we observe the Aboav-Weaire Law, in which low edge count cells have neighbours with a higher edge count, and vise versa, which can be approximated by the formula $m_n = 5 + 2 / n$ (Fig.~\ref{fig:3}f). Taken together, this suggests that for low $p_0$ there is more variability in cell sizes and thus pressure, and small cells tend to neighbour larger cells, creating higher curvature for low $p_0$. For high $p_0$, cells have zero tension and are floppy, resulting in very high curvatures for most edges.

For low $p_0$, there are larger differences in size for cells with different neighbour counts, meaning there are larger pressure differences for low $p_0$ than high $p_0$, while the tension scales like the square root in area. (do a simple scaling? Is it like size squared vs size? Show as 3rd figure?) Moreover, there is more variation in cell neighbour count for low $p_0$, which would further increase the mean curvature, as one would see larger differences in relative pressures between cells (Fig.~\ref{fig:3}e). In contrast, as $p_0$ increases, more cells attain their target area and perimeter, and thus are under no pressure and tension, and so the curvature is degenerate and can take larger values.

\begin{figure}[h!]
\includegraphics[width=\columnwidth]{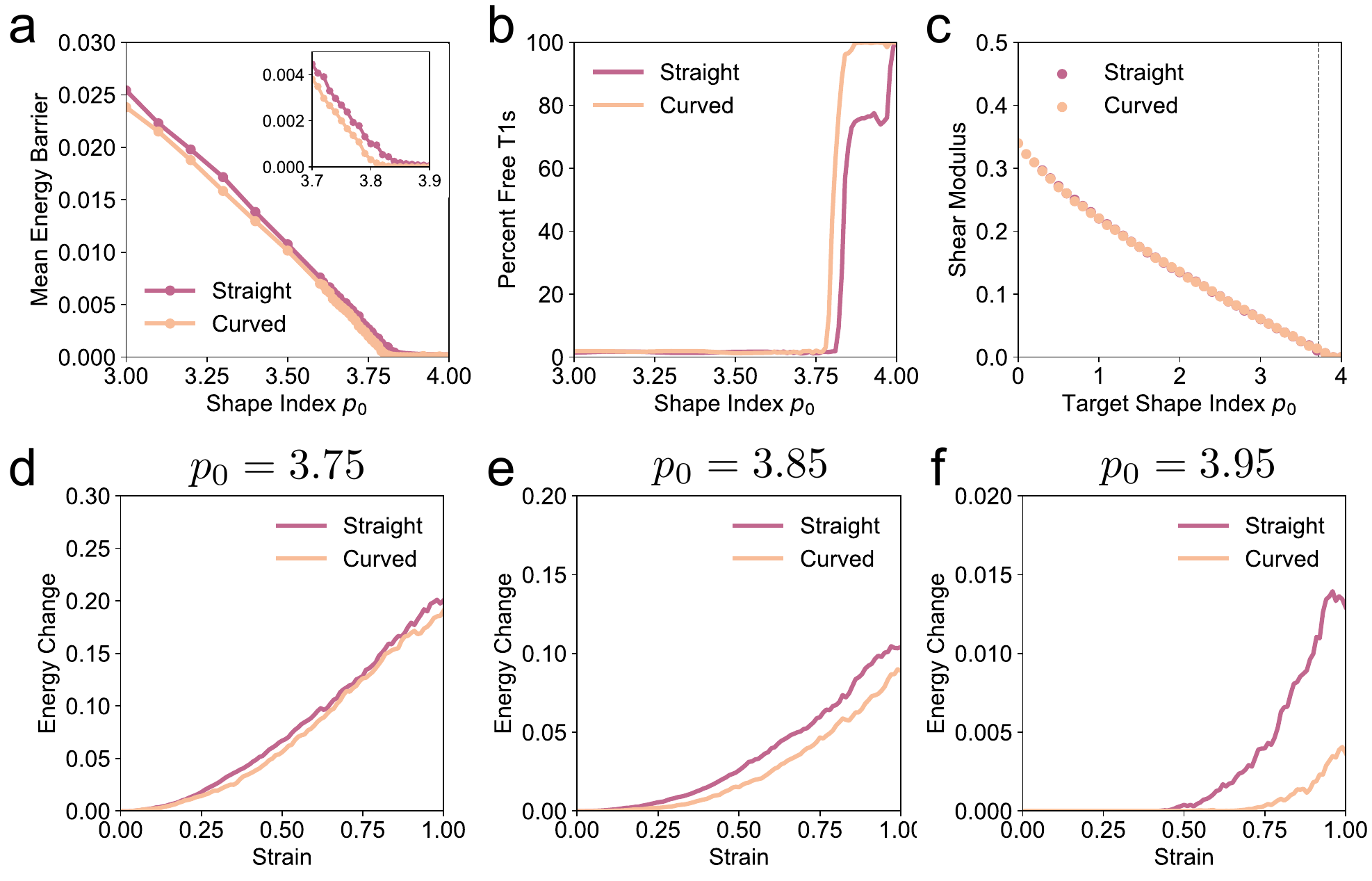}
\caption{Curved edges maintain fluidity over large strains. (a) Mean energy barrier during a T1 transition against $p_0$, for curved and straight edge vertex models. (b) Proportion of T1 transitions with no energy barrier. (c) Linear shear modulus against target shape index $p_0$. The dashed line shows $p_0 = p_6 \approx 3.72$. Averaged over 10 simulations. (d) Mean energy change against shear strain, for $p_0 = 3.75$, (e), $p_0 = 3.85$, and (f) $p_0 = 3.95$. Averaged over 100 simulations.}
\label{fig:4}
\end{figure}

Finally, we study the mechanical response of a disordered tissue using both curved and straight edges. To measure the energy barrier for a T1 transition, we calculate the energy required to shrink an edge to zero length and take the average across all edges and multiple simulations. As before, we observe a consistently lower energy barrier in the curved vertex model (Fig.~\ref{fig:4}a). The mean energy barrier decreases as $p_0$ increases, and becomes almost zero sooner for the curved vertex model, at around $p_0 = 3.80$, while the straight vertex model becomes almost zero at $p_0 = 3.85$. Moreover, the percent of T1s that have zero energy barrier is higher in the curved-edge vertex model (Fig.~\ref{fig:4}b). Additionally, for the straight-edge vertex model, the percent of zero energy T1s increases in steps. The second increase likely occurs because cells with 5-sides must become squares during a T1 transition, thus we need $p_0 > 4$ for them to have zero energy barrier, which is a qualitative difference between the curved-edge and straight-edge models. In contrast, the curved-edge model can bulge the edges outwards to account for this change in edge count.

We also measured the shear of each tissue. We start from a disordered configuration and apply a simple shear transformation to both the vertex positions and periodic boundaries, such that $x -> x + \epsilon y$, where $\epsilon$ is the strain, which increases from $\epsilon = 0$ to $\epsilon = 1$. For each value of strain we minimise energy by allowing vertices to move but not the boundaries.

Using $\epsilon = 0.01$ we calculate the linear shear modulus by estimating the second derivative of energy with respect to strain as
\begin{equation}
    G = \frac{2 \delta E}{\epsilon^2 A}
\end{equation}
where $\delta E$ is the difference in energy before and after shear and $A = L_x L_y$ is the area of the tissue. For all values of $p_0$, we find no significant different between the curved and straight vertex models (Fig.~\ref{fig:4}c). In both cases, the shear modulus decreases as $p_0$ increases and is zero around $p_0 = p_6$. Such behaviour is not unexpected. The linear shear only applies a small strain to the cell shapes, thus curved edges are unlikely to change the mechanical response as a result.

However, as we apply larger strains cells rearrange at lower strains, on average, in the curved-edge vertex model, resulting in a lower energy for a given strain when the strain is large (Fig.~\ref{fig:4}d-f). For $p_0 = 3.75$ the tissue jams under moderate shear strains, but has a small but consistently reduction in energy when using curved edges (Fig.~\ref{fig:4}d). As we enter the fluid regime at $p_0 = 3.85$ we see zero energy under small strains, but the tissue jams sooner for the straight-edge VM. Similarly, for $p_0 = 3.95$, the tissue remains fluid up to shear strains of 50\% for the straight-edge model, but around 75\% for the curved-edge model (Fig.~\ref{fig:4}f), likely due to the ability for pentagonal cells to rearrange at zero energy cost when edges may curve but not when they are straight.


In this manuscript, we have demonstrated how allowing for curved edges, in the form of circular arcs as per the Young-Laplace law, shifts the solid-to-fluid transition point in the vertex model by almost a whole polygon class, from $p_0 = 3.814$ to $p_0 = 3.729$ (Fig.~\ref{fig:2}). Within the solid phase, curved edges are important for determining the cell shape. Consistent with Lewis's law, cells with smaller neighbours are smaller on average, and, consistent with the Aboav-Weaire law, neighbour larger cells which themselves have more neighbours, resulting in small cells bulging outwards, while larger cells curve inwards (Fig.~\ref{fig:3}). This effect decreases as the target shape index $p_0$ increases until cells begin to achieve both their preferred area and shape index, becoming floppy with highly curved interfaces. Within the fluid phase, we find that curved-edges allow tissues to be deformed for zero energy for larger shear strains compared to the straight-edge model (Fig.~\ref{fig:3}). As well as quantitatively changing the mechanics of the vertex model, there is the qualitative difference in that lower polygon class cells may rearrange for no energy cost by bulging outwards to maintain their perimeter, allowing more of the cells to rearrange compared to in the straight-edge model.



Overall, this work highlights how curved edges help cells reduce stress and allow the tissue to fluidise at lower target shape index and for larger strains, than when edges are forced to be straight. Not only does employing curved edges more closely match the outward features of many realistic biological systems, but ignoring the edge-curvature degree of freedom can have significant consequences when studying the mechanics of biological processes like tissue shape change or invasion~\cite{perrone2016non}. Here, we have limited ourselves to a tissue with homogeneous cells and at equilibrium. For cells with different preferred sizes, such as during  division~\cite{ishimoto2014bubbly}, or tissues with activity, such as motility or tension fluctuations~\cite{schaumann2018force}, one would expect the impact of curved edges to be even more pronounced.

\bibliography{refs}
\end{document}